\documentclass[preprint,12pt,3p]{elsarticle}

\usepackage{amssymb}
\usepackage{cleveref}
\usepackage{float}
\usepackage{subfigure}
\usepackage{textgreek}
\usepackage{cleveref}
\usepackage{amsmath}
\usepackage{bm}
\usepackage{siunitx}
\usepackage{upgreek}
\usepackage{bigints}
\usepackage[nomarkers]{endfloat}

\biboptions{numbers,sort&compress}
\begin{document}

\begin{frontmatter}

\title{Synthesizing Cu-Sn nanowires alloy in highly-ordered Aluminum Oxide templates by using electrodeposition method}

\cortext[cor1]{Corresponding author: Mastooreh Seyedi
Department of Materials Science and Engineering, Clemson University, 161 Sirrine Hall, Clemson, SC 29634, USA
Email: sseyedi@clemson.edu Phone: 1-864-624-2838}

\author[label1]{Mastooreh Seyedi\corref{cor1}}
\address[label1]{Department of Materials Science and Engineering, Clemson University, 161 Sirrine Hall, Clemson, SC 29634, USA}

\author[label2]{Mozhdeh Saba}
\address[label2]{Department of Polymer Engineering and Color Technology, Amirkabir University of Technology, Tehran, Tehran, Iran}

\begin{abstract}
In this research a novel and simple electrochemical method is developed in order to facilitate the large-scale production of nanowires. The proposed electrochemical technique shows versatile controllability over chemical composition and crystalline structure of Cu-Sn nanowires. Another important factor, which could be controlled by using this method, is the order structure of nanowires more accurately in comparison to conventional synthesizing procedures.
As a result, the Cu-Sn nanowires as well as Aluminum Oxide templates synthesized by using the proposed electrochemical method are examined due to their morphology and chemical structure to find a relation between electrodeposition's solution chemistry and materials properties of Cu-Sn nanowires.
The results show that the proposed electrochemical method maintains a highly-ordered morphology as well as versatile controllability over chemical composition of nanowires, which could be used to optimize the procedure for industrial applications due to low cost and simple experimental setup.
\end{abstract}


\end{frontmatter}

% \linenumbers

\section{Introduction}
\label{sec1}

Tin (Sn) based anodes are rigorously studied in order to increase the Li-ion batteries' capacities \cite{doi:10.1021/acsaem.7b00242,TIAN201246,REN20072447,YOON2018193,doi:10.1002/adfm.200305165,doi:10.1021/acsami.6b03383,SENGUPTA2017290,POLAT2014585,ZHUO2013601,Jiang2012,YAMAMOTO2016275,B914650D,QIN2007948,HASSOUN20061336,FERRARA20111469,HUANG20092693,B823474D,doi:10.1002/adma.200602035,Mukaibo01102003,TAMURA20041949,doi:10.1021/acsnano.5b05823,doi:10.1002/adma.201603219}.
Although, large volumetric strains, during $\mathrm{Li^{+}}$ insertion/extraction, could lead to mechanical failure of anodes and reduce the cyclability of Lithium-ion batteries \cite{Wu2016,Whiteley01012016,C5TA00884K}.
As a result, alloying the anode material (i.e. Sn) with mechanically stable elements (e.g. Ni, Co, Cu, graphene, etc.) method is developed in order to achieve promising cycling performance and utilizing the higher capacity of Tin (Sn) simultaneously \cite{doi:10.1021/acsaem.7b00242,TIAN201246,REN20072447,doi:10.1021/acsami.6b03383,FERRARA20111469,B914650D}.
Furthermore, nanostructured version of these Tin based alloys (e.g. nanowires, nanoparticles, etc.) could stabilize the volumetric strains more effectively, because of their small volume changes in comparison to thin films \cite{doi:10.1021/nl4036498}.
Although, synthesizing the nanostructures in large volume for industrial applications is challenging due to their expensive and time consuming production methods, such as: surfactant based techniques or self-assembling of 0D nanostructures \cite{doi:10.1021/acsanm.8b00722,doi:10.1021/acsanm.8b00326,doi:10.1021/acsanm.7b00289}.
However, electrodeposition methods (e.g. direct or alternating current techniques) have shown more accurate controllability over materials morphology and chemical composition of nanostructures in comparison to conventional production techniques as well as their easier scalability for industrial applications \cite{doi:10.1021/jp308772b,doi:10.1080/17458080802570617,Liu:15}.
Electrodeposition of metallic alloys in the ordered self-assembled templates (e.g. $\mathrm{Al_{2}O_{3}}$, $\mathrm{TiO_{2}}$, porous polycarbonate, etc.) is developed primarily to control the morphology of nanostructures \cite{C0NR00206B,B603442J,doi:10.1021/la8037473,Heidari2015,Heidari2016}.
The main disadvantage of conventional direct current electrodeposition methods in order to synthesize nanostructures in self-assembled templates, is the practical difficulties of producing a conductive self-assembled structure due to low conductivity of templates' materials \cite{doi:10.1021/jp809202h,Hekmat2014,doi:10.1021/la901521j,Gong2008,doi:10.1002/pssa.200461161,Belov2011,Shamaila2008,Wang2006,doi:10.1002/anie.200504025,Ahmad2016,doi:10.1002/adma.201003624,Vorobyova2016,Mebed2016,doi:10.1021/cg500324j}.
On the other hand, alternating current (AC) electrochemical deposition method is primarily developed to overcome the technical difficulties, related to removing the barrier layer of the anodic Aluminum Oxide (AAO) and coating this self-assembled template with conductive materials \cite{doi:10.1021/la060190c}.
As a result, the AC electrodeposition technique has a potential to facilitate the industrial scale production of nanostructures with a highly controlled morphology as well as their chemical structure \cite{B603442J}.
In this research, AC electrochemical deposition method has been deployed to synthesize highly-ordered Cu-Sn nanowires in Aluminum Oxide templates. In the first section of this research paper, two steps anodization technique, which is used to produce a highly-ordered template, as well as its morphological properties will be discussed in details. 
Then morphological properties as well as chemical structure of synthesized Cu-Sn nanowires will be investigated in order to examine their crystalline structures as well as controllability of nanowires' chemical composition.
All the experiments in this research are done at room temperature, which facilitates its generalization for industrial scale applications due to technical difficulties related to costly temperature controlling systems \cite{SCHIAVI2018711}.

\section{Materials and methods}
\label{sec2}

\subsection{Two step anodizing}
\label{sec2subsec1}

\begin{figure}[H]
 \centering
 \subfigure{\includegraphics[width=\textwidth]{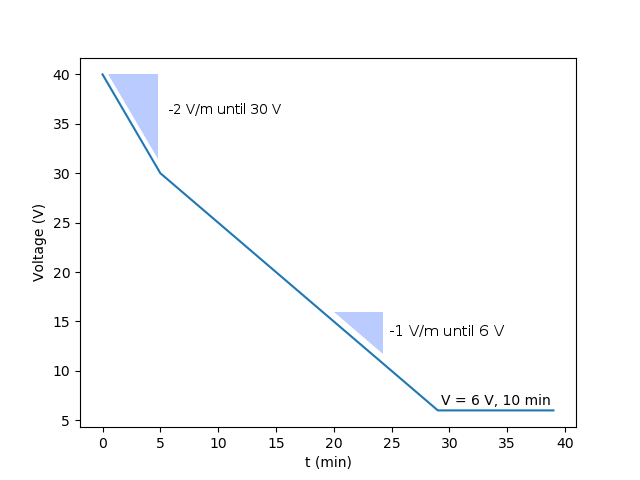}}
 \caption{Schematical representation of barrier layer thinning procedure, which shows decreasing the voltage to reduce the electrical impedance of Aluminum Oxide layer at the bottom of the pores.}
 \label{fig:1}
\end{figure}

Before anodizing of planar Aluminum samples (Merck KGaA, 99.95\%, 0.3 mm thickness - annealed), electrodes are electropolished ($\mathrm{A = 1 cm^{2}}$) in $\mathrm{HClO_{4}}$ 60 wt. \% solution.
In fact, the electropolishing voltage and temperature are fixed at 2 V and room temperature (\ang{25}C) respectively.
Also, the electropolishing time is optimized at 5min in order to achieve a highly smooth surface without primary amorphous Aluminum Oxide.
In both the first and second steps of anodization procedure, the electrodes are anodized in $\mathrm{C_{2}H_{2}O_{4}}$  0.3M solution.
Furthermore, anodization time, voltage, and temperature are fixed at 2h, 40 V and room temperature (\ang{25}C) respectively for first and second steps of this anodizing procedure.
Synthesized porous Aluminum Oxides after the first step of anodization are etched in a solution of 0.6M $\mathrm{H_{3}PO_{4}}$ 85 wt. \% - 0.2M $\mathrm{H_{2}CrO_{4}}$.
The etching temperature and time are fixed at \ang{60}C and 30min respectively.
The main challenge, in order to use Aluminum Oxide templates for electrodeposition purposes directly, is how to reduce the electrical resistance of $\mathrm{Al_{2}O_{3}}$ barrier layer, which prevents the electrical current flow through the thick insulative layer at the bottom of the pores (i.e. barrier layer) \cite{STEPNIOWSKI201580,C4CP04211E,doi:10.1021/nn800435n,STEPNIOWSKI201859,zhang_kielbasa_carroll_2009}.
In this research, the barrier layer thinning (BLT) procedure (i.e. reducing the second step anodization voltage gradually), which is shown as a voltage-time plot in \cref{fig:1} schematically, is used to reduce the electrical impedance of the Aluminum Oxide layer at the bottom of the pores.
Additionally, the electrical impedance of the electrodes, before and after BLT procedure, are examined by using impedance spectroscopy.

\subsection{AC nanowire electrodeposition}
\label{sec2subsec2}

The AC electrochemical deposition technique is employed to reduce $\mathrm{Cu^{2+}}$ and $\mathrm{Sn^{2+}}$ ions in the pores of self-assembled Aluminum Oxide template. In all the experiments of this section, pH as well as Boric acid ($\mathrm{H_{3}BO_{3}}$) concentration, root mean square (RMS) voltage, and AC signal's frequency are fixed at 1, 0.5M, 10 V, and 200 Hz respectively.
As a result, 10 samples are prepared to investigate the effect of Tin Sulfate ($\mathrm{SnSO_{4}}$) concentration on the chemical composition of produced nanowires. The $\mathrm{SnSO_{4}}$ concentration is changed from 0 to 0.5M and chemical composition as well as crystalline structure of deposited nanowires are examined by using energy dispersive spectroscopy (EDS) and X-ray diffraction respectively.

\subsection{Materials characterization}
\label{sec2subsec3}

Samples' characterization, which was used to investigate morphology of self-assembled templates as well as Cu-Sn nanowires, was done with a field emission scanning electron microscope (FE-SEM) Quanta 3D FEG (FEI, Phillips, The Netherlands).
In order to examine the pore sizes as well as morphology (i.e. order structure) of nanowires, ImageJ \cite{Rueden2017} image processing software is used to obtain quantitative information on the average diameter of the self-assembled templates' pores' diameter and length, as well as diameter of the electrodeposited Cu-Sn nanowires.
The crystalline structure of the Cu-Sn nanowire alloy was analyzed by X-ray diffraction using a Rigaku Ultima IV diffractometer with Co K radiation and operating parameters of 40 mA and 40 kV with a scanning speed of \ang{1} per minute and step size of \ang{0.02}.
Finally, the impedance spectroscopy of the anodized samples were done by using a MultiPalmSens4 potentiostat in order to compare the electrical resistance of anodizied samples before and after barrier layer thinning procedure.

\section{Results and discussion}
\label{sec3}

\subsection{Aluminum Oxide morphology}
\label{sec3subsec1}

\begin{figure}[H]
 \centering
 \subfigure[Pore size distribution of AAO template.]{\label{fig:2a}\includegraphics[width=0.45\textwidth]{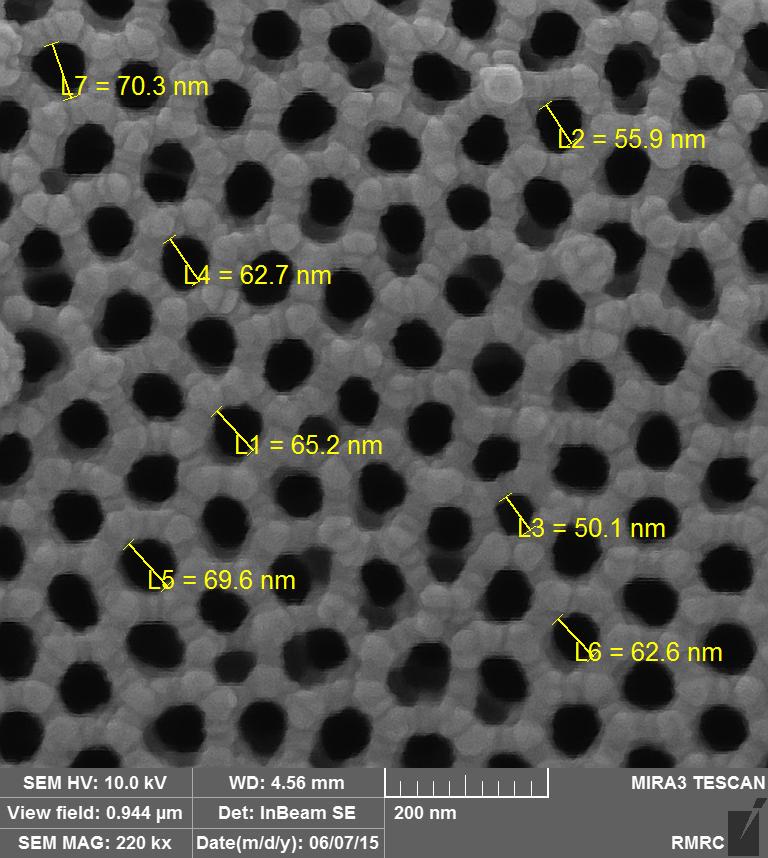}}\hspace{5mm}
 \subfigure[FFT result of AAO template, which shows perfect honeycomb structure due to 6 bright dots in its FFT spectrum.]{\label{fig:2b}\includegraphics[width=0.45\textwidth]{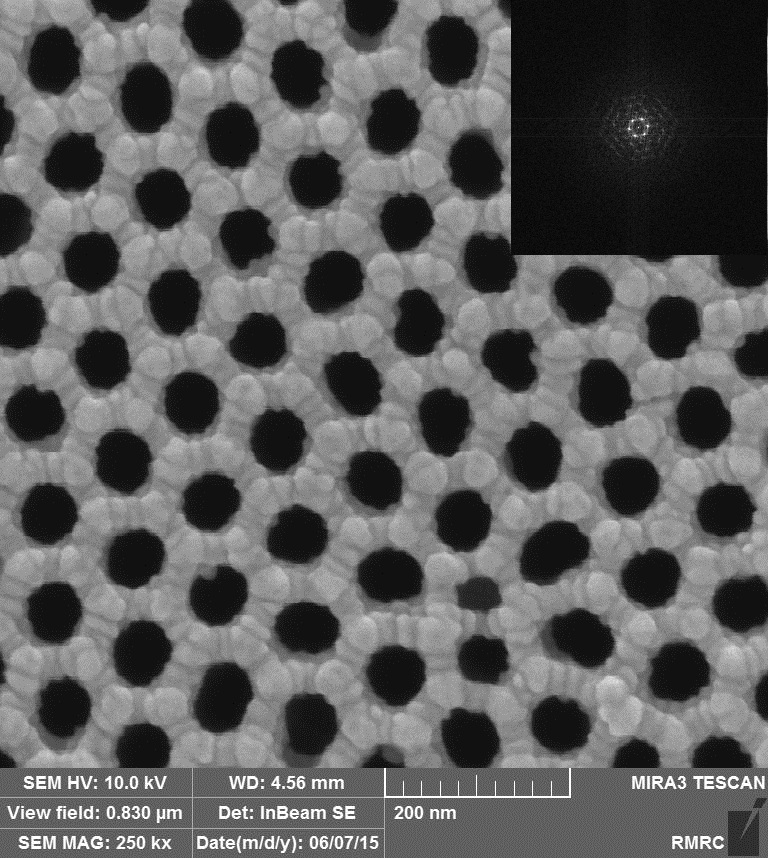}}\par
 \medskip
 \caption{Pore size distribution and FFT result of AAO template after two step anodization procedure.}
 \label{fig:2}
\end{figure}

\begin{figure}[H]
 \centering
 \subfigure[Thickness estimation of AAO template.]{\label{fig:3a}\includegraphics[width=0.45\textwidth]{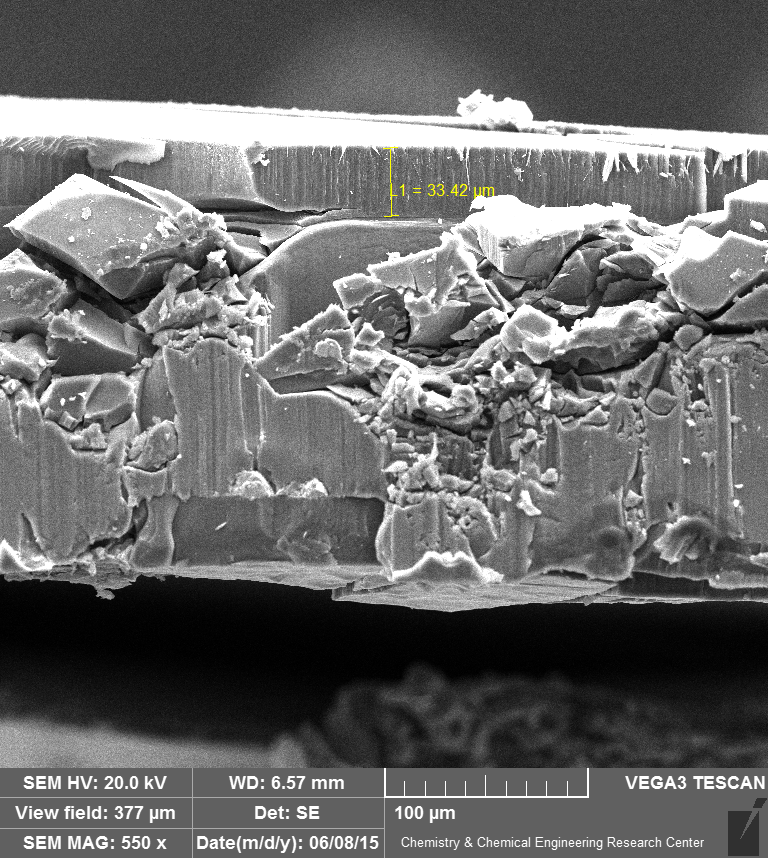}}\hspace{5mm}
 \subfigure[Pore wall estimation of AAO template.]{\label{fig:3b}\includegraphics[width=0.45\textwidth]{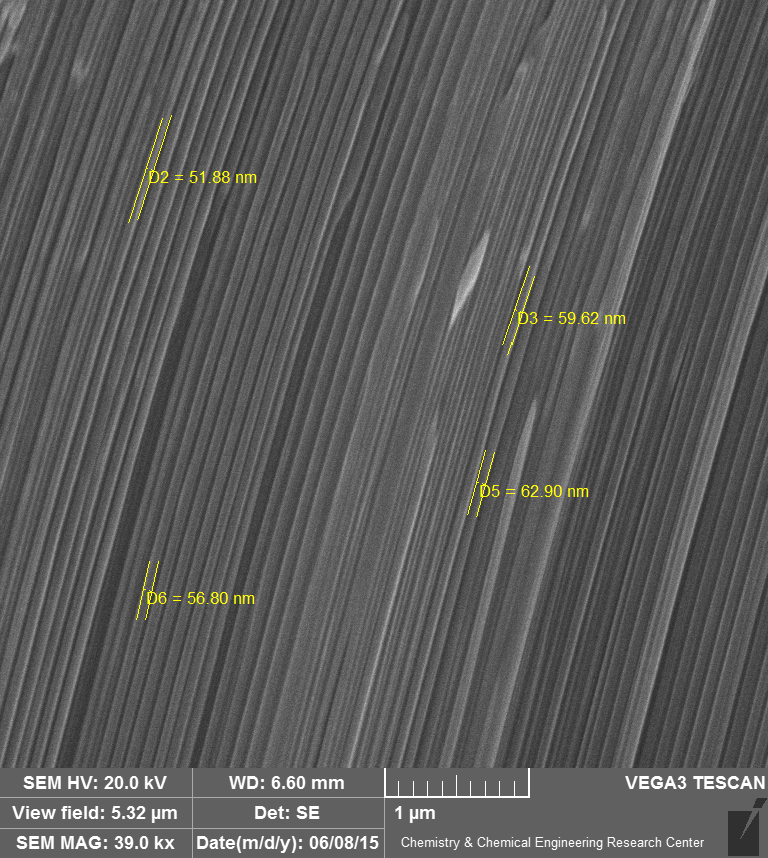}}\par
 \medskip
 \caption{Thickness and pore wall estimation of AAO template, which could be used to calculate the aspect ratio of self-assembled pores.}
 \label{fig:3}
\end{figure}

The porous morphology of two steps anodized Aluminum Oxide is shown in \cref{fig:2}. According to \cref{fig:2}\subref{fig:2a}, which is analyzed by using image processing techniques, it could be understood that the average pore size of this self-assembled template is 60 nm.
Furthermore, the order structure of this porous medium is analyzed by using the fast Fourier transform (FFT) technique in order to examine the spatial structure of pores and their deviation from honeycomb structure.
As a result, according to \cref{fig:2}\subref{fig:2b}, the FFT result of this AAO microstructure shows 6 strong bright dots, which indicates that a perfect honeycomb structure is achieved after the second step of anodization.
Additionally, in order to examine the aspect ratio of self-assembled pores of AAO, a cross-sectional FE-SEM microscopy is done to estimate the thickness of Aluminum Oxide after the second step of anodization (cref. \cref{fig:3}).
As shown in \cref{fig:3}\subref{fig:3a}, the thickness of AAO template is about 33 $\bm{\mu}$m. As a result, the aspect ratio of pores, which is defined as the ratio of thickness over diameter, could be estimated as 550. Also, this aspect ratio will be increased for nanowires after AAO dissolution because of nanowires' radial shrinkage due to compressive residual stresses \cite{Tasaltin2010,Gorisse2013}.
The wall thickness of pores in self-assembled AAO template is estimated as 60 nm, which is shown in \cref{fig:3}\subref{fig:3b}.
This highly ordered structure after the second step of anodization is achieved because: the quantum dots are created on the electrode's surface after etching step, which could facilitate the directional growth of AAO as well as controlling of its diameter more precisely \cite{doi:10.1063/1.1834987}.

\subsection{Impedance spectroscopy of Aluminum Oxide template's barrier layer}
\label{sec3subsec2}

\begin{figure}[H]
 \centering
 \subfigure[Impedance magnitude versus frequency for before and after BLT procedure.]{\label{fig:4a}\includegraphics[width=0.45\textwidth]{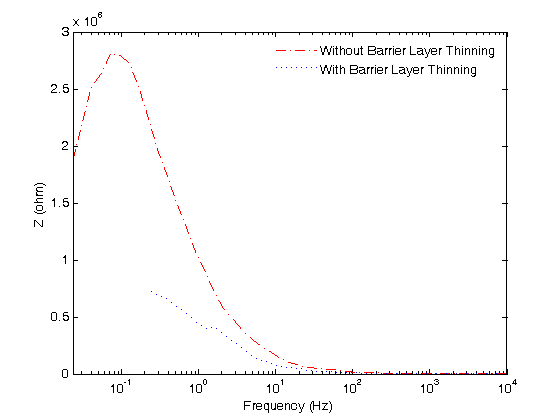}}\hspace{5mm}
 \subfigure[Imaginary part versu real part of impedance plot (Nyquist plot) for before and after BLT procedure.]{\label{fig:4b}\includegraphics[width=0.45\textwidth]{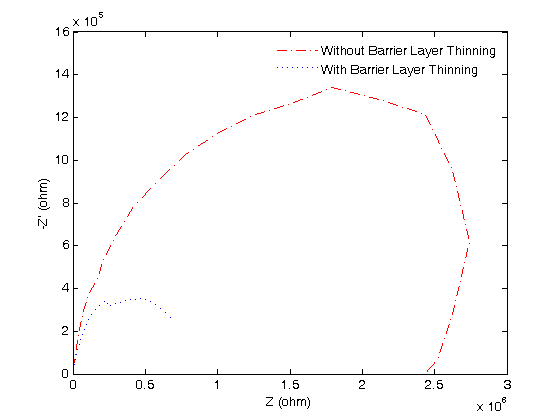}}\par
 \medskip
 \caption{Effect of BLT procedure on Nyquist plots of AAO template.}
 \label{fig:4}
\end{figure}

\begin{figure}[H]
 \centering
 \subfigure[Electrical circuit of Nyquist plot before BLT procedure.]{\label{fig:5a}\includegraphics[height=0.2\textwidth]{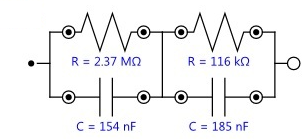}}\hspace{5mm}
 \subfigure[Electrical circuit of Nyquist plot after BLT procedure.]{\label{fig:5b}\includegraphics[height=0.2\textwidth]{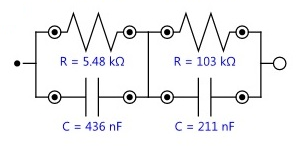}}\par
 \medskip
 \caption{Electrical circuits equivalence of Nyquist plots before and after BLT procedure.}
 \label{fig:5}
\end{figure}

Impedance spectroscopy is done in order to examine the electrical resistance of barrier layer before and after the thinning procedure.
Additionally, the impedance of barrier layer directly could be related to its thickness as \cite{Sulka2007,BOUCHAMA2013676}:
\begin{align}
 & Z = \frac{1}{(j \omega)^{a} C_{bl}} \label{eq1} \\
 & d_{bl} = \frac{\epsilon_{r} \epsilon_{0} S}{C_{bl}} \label{eq2}
\end{align}
Where $Z$ is the electrical impedance, $j = \sqrt{-1}$ is the imaginary unit, $\omega$ is the frequency, $a$ is frequency scattering factor, $C_{bl}$ is the barrier layer capacity, $d_{bl}$ is the barrier layer thickness, $\epsilon_{r}$ is the relative electrical permittivity, $\epsilon_{0}$ is the electrical permittivity of vacuum, and $S$ is the surface area of the sample.
The impedance magnitude versus frequency and its real part versus imaginary part (Nyquist plot) for before and after the barrier layer thinning procedure are shown in \cref{fig:4}\subref{fig:4a} and \cref{fig:4}\subref{fig:4b} respectively.
The equations \ref{eq1} and \ref{eq2} are used to fit them into the Nyquist plots (cref. \cref{fig:4}) and as a result, the obtained values for barrier layer thicknesses before and after BLT procedure are 20nm and 5nm respectively.
This calculation shows that the BLT procedure reduced the barrier layer thickness 4 times smaller, which could increase its conductivity and facilitate the AC electrochemical deposition step.
Additionally, the electrical circuits equivalence of the Nyquist plots are extracted due to the fitted parameters which are shown in \cref{fig:5}.
According to these electrical circuits, it could be understood that the second resistance/capacitance pair remained constant before and after BLT procedure. However, the electrical resistance of first resistance/capacitance pair is reduced by three orders of magnitude, which shows the electrical resistance is reduced after BLT procedure significantly (cref. \cref{fig:5}).

\subsection{Cu-Sn nanowires}
\label{sec3subsec3}

\begin{figure}[H]
 \centering
 \subfigure[Long-range order structure of Cu-Sn nanowires after AAO template dissolution.]{\label{fig:6a}\includegraphics[width=0.45\textwidth]{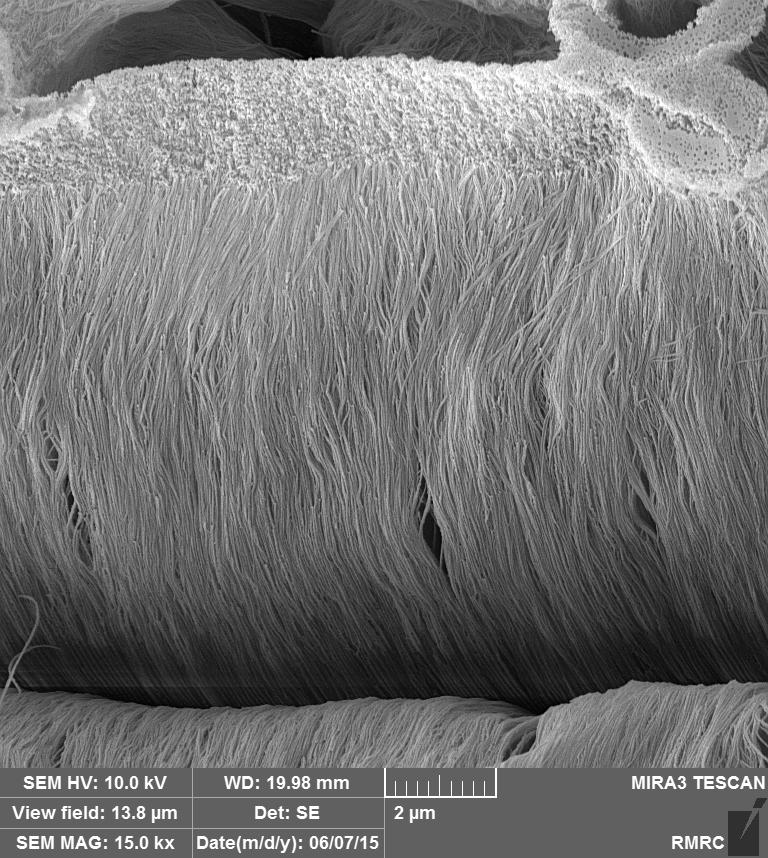}}\hspace{5mm}
 \subfigure[Diameter estimation of Cu-Sn nanowires in order to estimate their aspect ratio.]{\label{fig:6b}\includegraphics[width=0.45\textwidth]{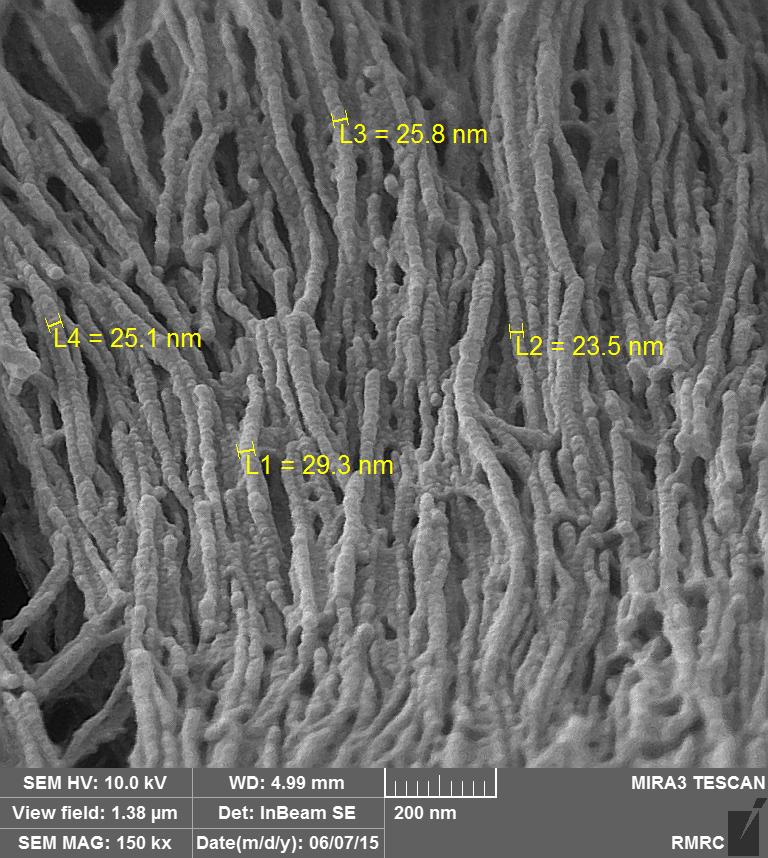}}\par
 \medskip
 \caption{FE-SEM microscopy images of Cu-Sn nanowires after AAO template dissolution.}
 \label{fig:6}
\end{figure}

\begin{figure}[H]
 \centering
 \subfigure[EDS spectrum of Cu-Sn nanowires, which shows their chemical composition.]{\label{fig:7a}\includegraphics[width=0.8\textwidth]{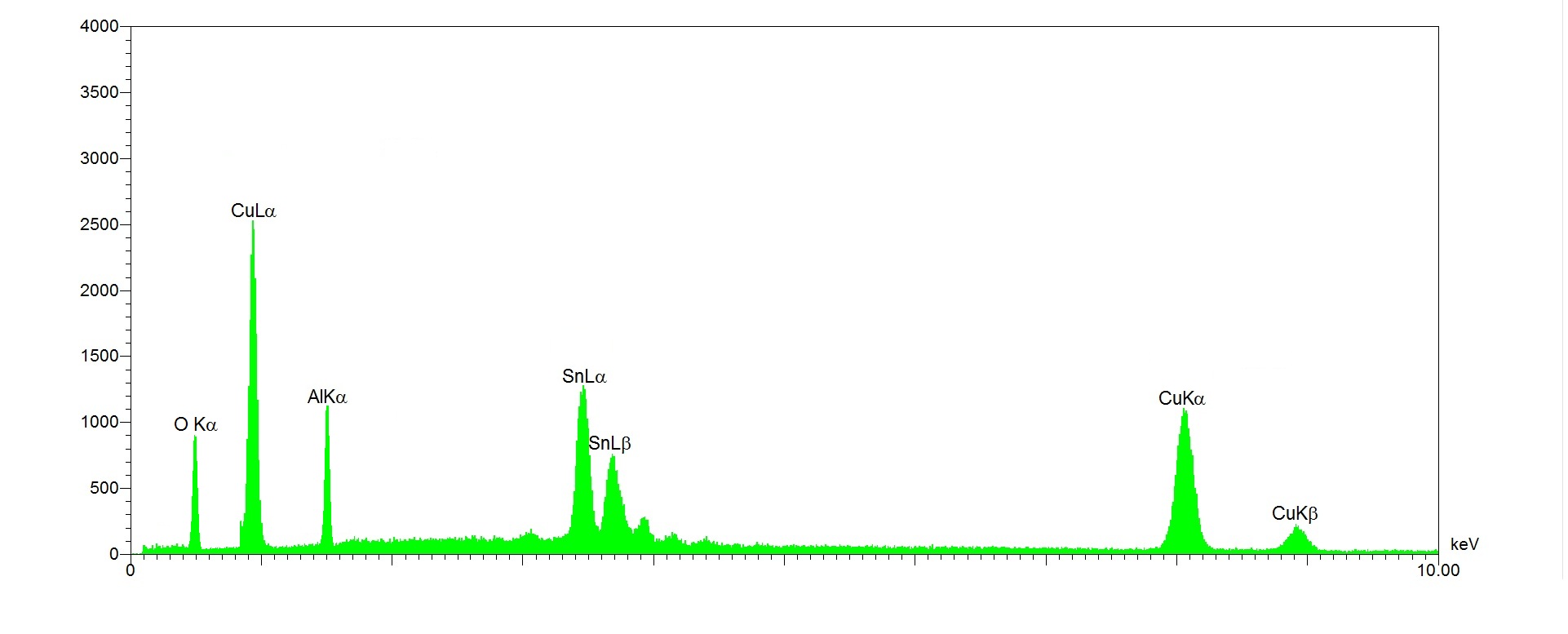}}\hspace{5mm}\par
 \medskip
 \subfigure[X-ray diffraction spectrum of Cu-Sn nanowires as well as their crystalline structures.]{\label{fig:7b}\includegraphics[width=0.8\textwidth]{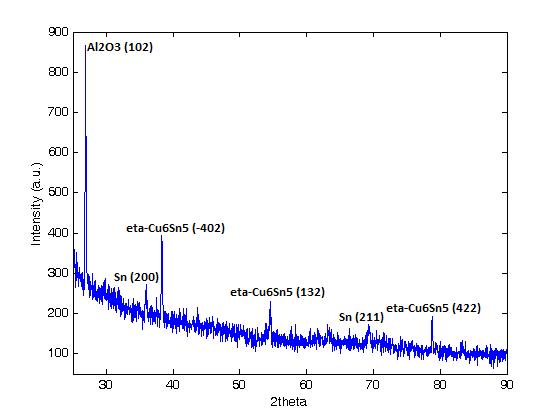}}\par
 \medskip
 \caption{EDS and X-ray diffraction spectra of Cu-Sn nanowires.}
 \label{fig:7}
\end{figure}

\begin{figure}[H]
 \centering
 \subfigure{\includegraphics[width=\textwidth]{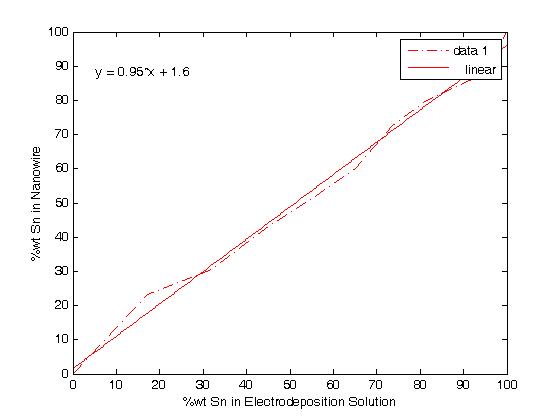}}
 \caption{Experimental data and linear fitted equation for chemical compositions of solution and nanowires of 10 samples.}
 \label{fig:8}
\end{figure}

FE-SEM microscopy technique is used to investigate the morphology of Cu-Sn nanowires after dissolution of AAO template in 1M NaOH solution. In \cref{fig:6}, Cu-Sn nanowires are shown in two different resolutions, which show their long-range order structure (cref. \cref{fig:6}\subref{fig:6a}) as well as the diameter of the nanowires (cref. \cref{fig:6}\subref{fig:6b}, 25 nm). As a result, according to \cref{fig:6}\subref{fig:6b}, it could be understood that the aspect ratio of the nanowires are increased by a factor of 2, which could increase their surface to volume ratio as well as their chemical reactivity for practical applications.
Also, the EDS and X-ray diffraction techniques are used to examine the chemical composition and crystalline structure of Cu-Sn nanowires respectively (cref. \cref{fig:7}).
According to \cref{fig:7}\subref{fig:7a}, there are some residual $\mathrm{Al_{2}O_{3}}$ due to presence of Aluminum and Oxygen peaks. These residual Aluminum Oxide could be eliminated by increasing the AAO template dissolution time.
Additionally, due to \cref{fig:7}\subref{fig:7b}, it is shown that the cystalline structure of nanowires is a mixture of unreacted crystalline Sn and $\bm{\eta}\mathrm{-Cu_{6}Sn_{5}}$ intermetallic compound.
Due to conventional DC electrochemical deposition procedures, the growth of intermetallic compounds needs a post heat treatment step to facilitate the re-nucleation and growth of crystalline structures \cite{5424866,5991646,Marquez2011,s16040431,doi:10.1080/17458080.2010.538442,C4TC02428A,doi:10.1111/j.1744-7402.2009.02364.x,Shi2014,Carlier01052006}.
As a result, the observed intermetallic compound ($\bm{\eta}\mathrm{-Cu_{6}Sn_{5}}$, cref. \cref{fig:7}\subref{fig:7b}) in Cu-Sn nanowires could be justified due to increasing the temperature during AC electrochemical deposition because of the high resistivity of pores' walls.
In fact, the electrical current flow chose the low resistance pathway through the barrier layer in order to reduce Copper and Tin ions, but the excess amount of electrical current flow through pore wall pathway will generate local thermal energy ($e = \frac{|\mathbf{J}|^{2}}{\sigma}$. where $e$ is the thermal energy, $\mathbf{J}$ is the electrical current density vector, and $\sigma$ is the electrical conductivity), which causes local heat treatment of the amorphous mixture of Copper and Tin \cite{Jang_electrochemicalproperties}.
Furthermore, $\bm{\eta}\mathrm{-Cu_{6}Sn_{5}}$ intermetallic compound shows promising higher efficiency in Lithium-ion batteries \cite{ORTIZ2014331}.
As a result, this AC electrochemical deposition technique could be optimized to eliminate the unreacted Sn in the nanowires by controlling the voltage, frequency, and chemical composition of the solution.
Additionally, by using EDS results, it is possible to find a direct relation between the solution chemistry and the nanowires chemical composition, which could be used to optimize this AC electrochemical deposition procedure aiming to maximize the amount of $\bm{\eta}\mathrm{-Cu_{6}Sn_{5}}$ intermetallic compound and efficiency of Lithium-ion batteries.
Hence, the solution and nanowires chemical compositions of 10 samples (swept $\mathrm{SnSO_{4}}$ concentration from 0 to 0.5M) are tabulated in \cref{table:1}.
\begin{center}
\begin{table}[H]
\centering
\tabcolsep=0.0005cm
\scalebox{1}{ 
 \begin{tabular}{||c c c||}
 \hline
 Sample No. & Sn wt. \% in solution & Sn wt. \% in nanowires \\ [0.5ex]
 \hline\hline
 1 & 0 & 0 \\
 \hline
 2 & 17.19 & 23.21 \\
 \hline
 3 & 31.84 & 30.69 \\
 \hline
 4 & 44.46 & 42.52 \\
 \hline
 5 & 55.46 & 51.66 \\
 \hline
 6 & 65.13 & 60.03 \\
 \hline
 7 & 73.70 & 72.76 \\
 \hline
 8 & 81.34 & 80.05 \\
 \hline
 9 & 94.39 & 87.58 \\
 \hline
 10 & 100 & 100 \\ [1ex]
\end{tabular}}
 \caption{Chemical compositions of solution and nanowires obtained from EDS.}
 \label{table:1}
\end{table}
\end{center}
The experimental data as well as the linear fitted equation for the relation of the solution and nanowires chemical compositions are plotted in \cref{fig:8}.
As a result, the final fitted equation which could relate the chemical composition of solution with nanowires structure, is obtained as:
\begin{align}
 wt. Sn \in NWs = 0.95 \times wt. Sn \in Solution + 1.6 \label{eq3} 
\end{align}
This relation (cref. equation \ref{eq3}) could be used to optimize the chemical composition as well as crystalline structure of Cu-Sn nanowires and ultimately increase the efficiency of Lithium-ion batteries.

\section{Conclusions}
\label{sec4}

In this research, an AC electrochemical deposition technique is developed, which could be easily used to synthesize metallic nanowires with highly-ordered structure and reasonable controllability over the chemical composition as well as the crystalline structure.
Additionally, this technique could facilitate the large-scale production of nanostructures due to the in-situ heat treatment of nanowires as well as the room temperature operating environment.
As a result, this technique could be generalized to develop industrial scale coating facilities, which could be used in Lithium-ion battery production as well as other industries, such as: biomedical applications \cite{AMIRJANI2014210,Bressloff2016,doi:10.1080/10255842.2017.1286650,doi:10.1002/mawe.201700241}, oil and gas extraction plants \cite{YOUSEFI2015138,GHALAMBAZ201752,TAIE20185124}, and nanoparticles technologies \cite{Yousefi2015}.
Finally, these Cu-Sn nanowires' production technique should be optimized by using the proposed electrochemical synthesizing procedure, and be examined in assembled Lithium-ion batteries to accurately measure their capacity as well as efficiency in order to achieve higher cyclability.

%% References


\clearpage

\bibliographystyle{elsarticle-num}

\bibliography{bibliography}

\end{document}